\documentstyle[11pt,aaspp4,tighten]{article}

\tightenlines

\begin{document}

\title{X-ray Observations of the  Broad-Line Radio Galaxy 3C~390.3}

\author{Karen M. Leighly\altaffilmark{1,2}\affil{leighly@postman.riken.go.jp} 
Paul T. O'Brien\altaffilmark{3}\affil{pto@star.le.ac.uk} 
Rick Edelson\altaffilmark{4}\affil{edelson@spacly.physics.uiowa.edu} 
Ian M.  George\altaffilmark{5,6}\affil{george@lheavx.gsfc.nasa.gov}
Matthew A. Malkan\altaffilmark{7}\affil{malkan@bonnie.astro.ucla.edu}
Masaru Matsuoka\altaffilmark{1}\affil{matsuoka@postman.riken.go.jp}
Richard F. Mushotzky\altaffilmark{5}\affil{mushotzky@lheavx.gsfc.nasa.gov}
Bradley M. Peterson\altaffilmark{8}\affil{peterson@payne.mps.ohio-state.edu}} 
\altaffiltext{1}{Cosmic Radiation Laboratory, RIKEN, Hirosawa 2--1, Wako-shi, 
Saitama 351, Japan}
\altaffiltext{2}{Current address: Columbia Astrophysics Laboratory,
538 West 120th Street, New York, NY 10027}
\altaffiltext{3}{Department of Physics \& Astronomy, University of
Leicester, University Road, Leicester, LE1~7RH, U.K.}
\altaffiltext{4}{Department of Physics and Astronomy, University of Iowa, Iowa
City, IA 52242-1479}
\altaffiltext{5}{Code 660.2, NASA Goddard Space Flight Center, Greenbelt, MD
20771}
\altaffiltext{6}{Also Universities Space Research Association}
\altaffiltext{7}{Department of Astronomy, University of California, Los
Angeles, CA, 90024-1562}
\altaffiltext{8}{Department of Astronomy, Ohio State University, 174
W. 18th Ave., Columbus, OH 43210-1106}

\slugcomment{Submitted to {\it The Astrophysical Journal}}


\begin{abstract}

We present the data and preliminary analysis for a series of 90 {\it
ROSAT} HRI and two {\it ASCA} observations of the broad-line radio
galaxy 3C~390.3. These data were obtained during the period 1995
January 2 to 1995 October 6 as part of an intensive multiwavelength
monitoring campaign. The soft X-ray flux in the {\it ROSAT} band
varied by nearly a factor of four during the campaign, and the
well-resolved light-curve shows several distinct features. Several
large amplitude flares were observed, including one in which the flux
increased by a factor of about 3 in 12 days. Periods of reduced
variability were also seen, including one nearly 30 days long.  While
the HRI hardness ratio decreased significantly, it is apparently
consistent with that expected due to the detector during the
monitoring period.

The two {\it ASCA} observations were made on 1995 January~15 and 1995
May~5.  The 0.5--10.0 keV spectra can be adequately described by an
absorbed power-law.  There is no evidence for a soft excess in the
{\it ASCA} spectra, indicating that the {\it ROSAT} HRI is sampling
variability of the X-ray power-law.  A broad iron line was observed in
a longer 1993 {\it ASCA} observation, and while there is statistical
evidence that the line is present in the 1995 spectra, it could
not be resolved clearly.  There is evidence, significant at $>90$\%
confidence, that the photon index changed from 1.7 to 1.82 while the
flux increased by 63\%.  The spectral change can be detected in the spectra
below 5 keV, indicating that the origin cannot be a change in ratio of
reflected to power-law flux.  A compilation of results from {\it ASCA}
and {\it Ginga} observations show that on long time scales the
intrinsic photon index is correlated with the flux.

\end{abstract}

\keywords{galaxies: individual (3C~390.3) -- X-rays: galaxies --
galaxies: active}

\section{Introduction}

3C~390.3 is a luminous ($L_{X(2-10)}\sim 2-4 \times 10^{44} \,\rm ergs
\, s^{-1}$) nearby (z=0.057) broad-line radio galaxy located in the
North Ecliptic cap.  It is well known as the prototypical source of
broad double-peaked and variable $H\beta$ lines (Eracleous \& Halpern
\markcite{7} 1994; Veilleux \& Zheng \markcite{21} 1991).  It has a
superluminal compact kiloparsec scale radio jet as well as two
extended radio lobes, each with a hot spot (e.g. Alef et al.\
\markcite{1} 1996).  3C~390.3 is variable in all wave bands on time
scales from days to years, including the optical (Barr et al.\
\markcite{2} 1980) and UV (Clavel
\& Wamsteker \markcite{3} 1987).  It is a bright and variable X-ray
source, and has been observed many times in X-rays (see Eracleous,
Halpern \& Livio \markcite{8} 1996 for a compilation).  Hard X-ray
variability by a factor of 2 was reported over a period of 6 weeks
when observed with {\it OSO 7} (Mushotzky, Baity \& 
Peterson \markcite{15} 1977).
Variability by a factor of 6 was observed among 6 observations made by
the EXOSAT observatory (Inda et al.\ \markcite{10} 1994).
Historically, the spectrum has generally been well described by an
absorbed power-law.  Evidence that the measured column density exceeds
the Galactic and may be variable has been found in several
observations (Eracleous et al. \markcite{8} 1996 and
references therein).  Photon index variability has also been found;
between two {\it Ginga} observations 2.5 years apart, Inda et al.\
\markcite{10} (1994) found a change in photon index from 1.76 to 1.54
when the flux decreased by $\sim 30$\%.  Evidence for an iron
$K\alpha$ line was also found in the {\it Ginga} spectra (Inda et al.\
\markcite{10} 1994).  The iron line was resolved in a 1993
{\it ASCA} observation and the measured velocity width of $\sim
16,500\rm\,km/s$ is consistent with an origin in an accretion disk at
about 250 gravitational radii (Eracleous et al.
\markcite{8} 1996).

During 1995, 3C~390.3 was subject to a monitoring campaign from radio
through X-rays.  Using the {\it ROSAT} HRI, we obtained the first
well-sampled X-ray light-curve on time scales from days to months.
This paper presents the {\it ROSAT} data as well as two {\it ASCA}
observations obtained during the monitoring period.  The soft X-ray
variability is briefly described, but the results of the detailed
timing analysis, and the results from the optical, UV and radio
monitoring will be reported elsewhere (Leighly
\& O'Brien \markcite{12} 1997, Dietrich et al.\
\markcite{5} 1997, O'Brien et al.\ \markcite{18} 1997, Leighly et al.\
\markcite{13} 1997).

\section{{\it ROSAT} Data} 

Ninety-three {\it ROSAT} HRI observations of 3C~390.3 each 1500
seconds with monitoring interval of three days were scheduled. Ninety
observations were successful, and only three observations were missed.
Excluding these, the mean interval was 3.22 days with standard
deviation 0.75 days. The first observation began 2 January 1995 5:05
and the last observation ended 6 October 1995 11:10.  The background
subtracted light-curve is shown in the upper panel of Figure~1, and
the {\it ROSAT} data sequence numbers, dates of observation, exposures
and count rates are given in Table~1.  Note that the dates are given
for the midpoint of the observation, and no barycentric correction has
been applied.

\begin{figure}[h]
\vbox to3.75in{\rule{0pt}{3.75in}}
\includegraphics{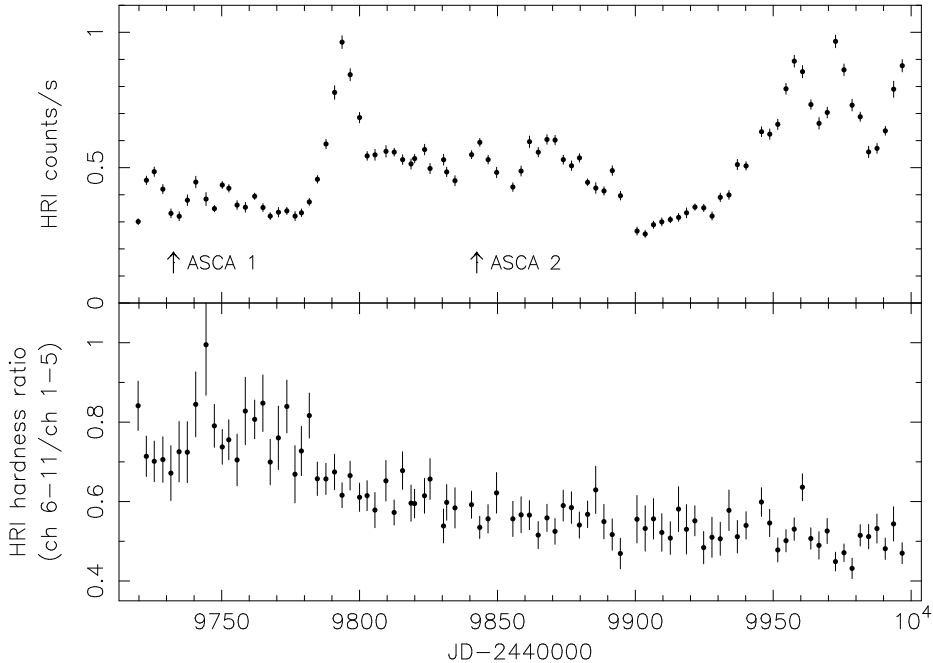}
\caption{Top: {\it ROSAT} HRI light-curve from monitoring observations of
3C~390.3.  The times of the two {\it ASCA} observations are marked
with arrows.  Using the distributed HRI response matrix and ignoring
the effects of the gain change, 1~count/s corresponds to 2.93 and
$3.04 \times 10^{-11}\rm\,ergs\,s^{-1}cm^{-2}$ for photon indices 1.7
and 1.82 respectively and $N_H=1.15\times 10^{21}\rm\, cm^{-2}$.
Bottom: The HRI hardness ratio shows that the spectrum becomes
steadily steeper as a function of time.  However, the steepening
likely is dominated by the temporal gain of the detector and cannot be
used to study the spectral variability.}
\end{figure}

The {\it ROSAT} HRI is an ideal instrument for monitoring the soft
X-ray flux from 0.1--2.0 keV.  The instrument is sensitive to low
fluxes because the background is low. The FWHM of the main component
of the image point spread function is $1.7^{\prime\prime}$, broadened
somewhat by the attitude reconstruction uncertainty of about
$6-10^{\prime\prime}$.  Thus the source photons can be extracted from
a small region, reducing further the statistical error from background
subtraction.  We used a source extraction region $1.5^{\prime}$ in
radius, and the background was estimated from an annulus with radii
$2.5^\prime$ and $5^\prime$.  These regions are near the center of the
detector where the image is fairly flat, so no off-axis response
correction as applied.   The background count rate was variable,
but because the background rate was only about 5\% of the source count
rate, it was only necessary to exclude intervals with very large
spikes. 

\section{{\it ROSAT} Analysis} 

\subsection{Variability}

The light-curve in Figure~1 shows that significant, large amplitude
soft X-ray variability occurred during the monitoring period.  We
chose the 3 day monitoring interval based on the fact that only low
amplitude variability was detected during single day observations by
{\it ROSAT} and {\it Ginga}, but measurable variability occurred
between observations several days apart.  This choice is vindicated by
the fact that significant variability was not found during individual
{\it ROSAT} observations (all of which were made within one or two
satellite orbits), but variability of up to 50\% was found between
adjacent observations.

The most noticeable feature in the {\it ROSAT} light-curve is a flare around
TJD~9800 in which the flux increased by a factor of 3 in about 12 days,
and then decreased by a factor of 2 in about 9 days.  A series of similar
flares is apparent after TJD~9950.  These contrast markedly with periods
of reduced variability, which seem to occur preferentially before and
after the flares and most notably during a period of about 30 days
starting at about TJD~9900.  A discussion and interpretation of this
distinctive variability is given in Leighly \& O'Brien \markcite{12}
(1997).

During the {\it ROSAT} campaign the average source count rate was
$0.5\rm\, cnt\,s^{-1}$ while the average background count rate in the
source region was $0.027\rm \,cnt\,s^{-1}$, yielding an average
uncertainty in the count rate of $0.017\,\rm cnt\,s^{-1}$.  The
average effective exposure was $\sim 1900\,\rm s$ and the average
signal-to-noise was $\sim 30$.  The ratio of the largest to smallest flux,
$R_{max}$, is 3.8, demonstrating that a large amplitude of variability
was observed in soft X-rays during this campaign.  The fractional
variability amplitude ($F_{var}$) is defined as the standard deviation
of the flux divided by the mean flux, and gives a measure of the
magnitude of the variability during the observation (e.g. Edelson et
al.\ \markcite{6} 1996).  The true standard deviation is used, i.e.
the mean measurement error has been subtracted in quadrature.  For
this series of observations, the $F_{var}$ is 0.328.  Because the
monitoring spanned a comparatively long period of time, the observed
amplitude of variability is large compared with previous observing
campaigns in X-rays. For example, in the previous AGN Watch monitoring
campaign of NGC 4151, the $F_{var}$ from 13 {\it ROSAT} PSPC and 4
{\it ASCA} observations in the 1--2 keV band was 23.9\%, but note that
the monitoring period was just 10 days.
 
The time scale of variability appears relatively long.  In comparison,
during an {\it ASCA} observation of MCG~--6-30-15, variability by a
factor of 1.5 was observed in 100 seconds (Reynolds et al.\
\markcite{19} 1995).  However, it has been found that various measures
of the variability time scale are correlated with the luminosity
(Lawrence \& Papadakis
\markcite{11} 1993).  In the {\it ASCA} observations reported here
(see below), the 2--10 keV luminosity was found to be $2-4\times
10^{44}\rm\, erg\,s^{-1}$.  Thus 3C~390.3 is a relatively luminous
object compared with many well studied X-ray variable Seyfert 1s
including MCG~--6-30-15 ($L_X=1.1\times 10^{43}\,\rm ergs\,s^{-1}$).
From the correlation between doubling time scale and luminosity
obtained from EXOSAT data by Lawrence \& Papadakis
\markcite{11} (1993), this luminosity should correspond to a doubling
time scale of $\sim 10^6\rm\,s$, which is approximately 12 days, or
the time scale of the flare.  Thus the longer observed time scale of
variability is consistent with the greater luminosity of this object.

\subsection{Spectral Variability}

The HRI hardness ratio, defined as the ratio of counts in channels
6--10 to 1--5, varies significantly during the monitoring period
(Figure 1).  While the spectral response of the {\it ROSAT} HRI is not
well determined, the hardness ratio gives some indication of
differences in spectra (David et al.\ \markcite{4} 1996).  However, it
is known that there is a temporal gain variation in the HRI, such that
the photons are detected in progressively lower channels, thus
producing smaller values of the hardness ratio (David et al.\
\markcite{4} 1996), as observed. This must be the source of at least
part of the hardness ratio change.  Two {\it ASCA} observations during
the monitoring period show significant spectral variability in the
same sense as the HRI hardness ratio change (see below), but repeated
spectral simulations using the HRI response matrix obtained before
launch and the ASCA best fit models give indistinguishable mean PHA
and hardness ratios.  While the shift in the
mean channel of the 3C 390.3 spectra is larger than the shift observed
from four calibration observations of the supernova remnant N132D
contemporaneous with the monitoring campaign (J. Pye, J.  Silverman
1996 p. comm.), that may be expected because the spectrum of 3C 390.3
is harder.  In conclusion, we can make no definite statement about the
spectral variability during the monitoring period from the HRI
hardness ratio.

\section{{\it ASCA} Data} 

Two 20ks {\it ASCA} observations were made during the {\it ROSAT}
monitoring campaign.  These started 1995 January 15 9:17 and 1995 May 5
23:57, and are hereafter referred to as the first and second observations.
The angles between the telescope axes (XRT) and the Sun were 101.3 and
79.2 degrees and the observing modes were 2CCD Bright mode and 1CCD
Faint mode for the first and second observations, respectively.  The
times of these observations are marked in the upper panel of Figure~1.
Events were excluded when the satellite was in the South Atlantic
Anomaly (SAA), and at the beginning of the observation when the
attitude was unstable. The following selection criteria were used:
coefficient of rigidity $>6\rm\,GeV/c$; elevation angle $>5^\circ$;
radiation belt monitor count rate (RBM\_CONT) $<200\rm\,cnts\,s^{-1}$.
For the SIS detectors, the hot and flickering pixels were removed, and
the further selection criteria were used: bright earth angle
$>15^\circ$; events collected within 4 readout cycles (32 and 16 s for
the first and second observations, respectively) after passage through
the SAA and the day--night terminator were excluded; events collected
above the threshold of 400 pixels per CCD per readout were excluded;
event grades 0 and 2--4 were used.  The standard pulse-height versus
rise time filter was applied to GIS events.  The source spectra were
obtained from regions of radius 4 arcmin and 6 arcmin for the SIS and
GIS respectively.  The background spectra were obtained from source
free regions of the same chip for the SIS, and from regions as close
as possible to the same distance from the optical axis as the source
region for the GIS.  SIS response matrices appropriate for the time of
the observation were made.  The spectra were fit
between 0.5--10.0 keV and 0.8--10.0 keV for the SIS and GIS
respectively.

The XRT--sun angle was somewhat low during the second observation, so
there was the possibility of contamination by light leakage. We
carefully examined the SIS spectra using various constraints on the
bright earth elevation angle and found no evidence for light leakage.

While most of the spectra were fit well by an absorbed power-law
model, residuals including a hard tail and a soft excess were observed
in fits of the SIS1 spectrum from the January observation.  In
addition, absorption column versus photon index $\chi^2$ contours from
the SIS0 and SIS1 spectra were inconsistent at the 90\% confidence
level.  This observation was made in 2CCD Bright mode, which means
that while data collected at medium bit rate were compressed on-board,
the greater telemetry rate during high bit rate transmission allowed
the data to be sent to the ground before compression.  The high bit
rate data must be compressed before analysis, but because that is done
on the ground, additional calibration including correction for echo
and dark frame error (DFE) is possible before compression.  During
this observation, approximately half the data were collected at high
bit rate.  Thus we collected corrected high bit rate spectra
separately and found that the contours generated from these spectra
from both detectors were consistent with the uncorrected SIS0 contour,
but the uncorrected SIS1 contour was offset.  Because of this
behavior, the medium bit rate data from SIS1 of the January
observation was judged anomalous, although the cause is not known, and
the high bit rate spectrum from that detector was used in further
spectral fits.  The observation log is given in Table~2.

\section{{\it ASCA} Analysis} 

The count rates listed in Table~2 indicate that the flux changed
significantly between the two observations.  The spectra from all four
detectors from both observations were fit simultaneously with a
power-law plus absorption model, allowing the normalization to be free
between the two observations.  The resulting fit was acceptable:
$\chi^2$ was 2262/2124 degrees of freedom (d.o.f.), the photon index
was 1.76 and the absorption column was $1.0\times
10^{21}\rm\,cm^{-2}$.  Allowing the photon index to vary between the
two observations gave a better fit with $\chi^2=2183$/2123~d.o.f.  The
change in $\chi^2$ of 79 indicates with $>99.9$\% confidence obtained using
the F test for one additional degree of freedom that a significant
change in photon index occurred between these two observations.
Alternatively, allowing only the absorption to vary yields
$\chi^2=2207$/2123~d.o.f.  Allowing both to vary gives
$\chi^2=2182$/2122~d.o.f., so there is no evidence that both the
absorption and the photon index vary.  When the absorption was equated
between the two observations, the best fit column was $1.1\times
10^{21}\rm
\,cm^{-2}$ and the two best fit indices were 1.68 and 1.80.  These
photon indices are in the range typical for an AGN, although flatter
than the average (e.g. Nandra \& Pounds \markcite{17} 1994).  
 
There is no evidence for a soft excess component in the residuals of
the power-law fit to the {\it ASCA} spectra.  The HRI response is
harder than that of the {\it ROSAT} PSPC.  There is a little
sensitivity to photons in the 0.1--0.4~keV range, but most of the
effective area is above 0.4~keV and therefore the band passes of the
HRI and {\it ASCA} SIS overlap almost completely.  This means that
the HRI is sampling the variability of the X-ray power-law.

A broad iron $\rm K\alpha$ line was discovered in the previous 1993
observation of 3C~390.3 (Eracleous et al. \markcite{8} 1996).  The
exposures here are shorter, and the presence of the broad line cannot
be clearly seen in the residuals (Figure~2).  A Gaussian line was
added to the model, with the parameters between the two observations
constrained to be equal.  When the energy is fixed at 6.4~keV in the
rest frame and the width is narrow, the fit was improved by
$\Delta\chi^2=-7$, significant with $>99$\% confidence evaluated using
the F~statistic.  Allowing either the line energy or the line width to
vary results in further improvement in fit by $\Delta\chi^2=-10$ which
is significant at $>99.5$\% confidence.  In this case, the line energy
shifts to a lower value, and the line width increases (Table 3), as
expected if the line is emitted from a rotating disk (e.g. Matt et al.
\markcite{26} 1992).  Finally, allowing the remaining parameter to
vary does not result in a significant improvement in fit
($\Delta\chi^2=1$).  This suggests that the broad line is certainly
present in the spectra, but the shorter exposures don't allow us to
clearly resolve it.  No reduction of $\chi^2$ results when the line
normalization is allowed to vary between the two observations.  The
parameters from the best fit spectra are listed in Table~3.  The
measured line parameters are consistent with those obtained by
Eracleous et al. \markcite{8} (1996).  The physical width of the line
was frozen at the best fit value to avoid fitting instability during
error analysis.  The quoted uncertainties are 90\% for two parameters
of interest ($\Delta\chi^2=4.61$).  Figure~3 shows the photon index vs
$N_H$ $\chi^2$ contours for the two observations (solid line).  These
indicate that the photon index varied at $>90$\% significance between
the two observations, while the absorption remained the same.

\begin{figure}[h]
\vbox to3.0in{\rule{0pt}{3.0in}}
\includegraphics{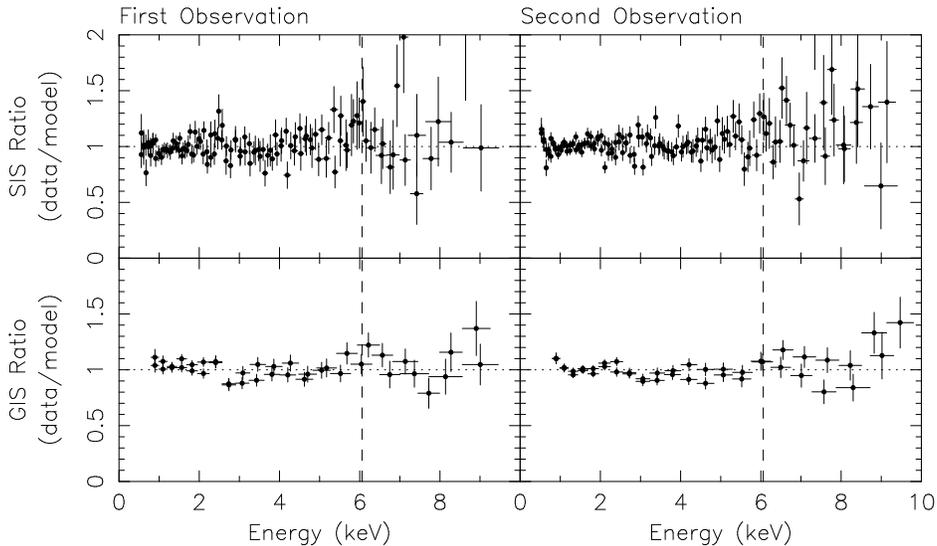}
\caption{The ratio of data to an absorbed power-law model for the
two {\it ASCA} observations.  The vertical dotted line marks the
location of the cosmologically redshifted iron $K\alpha$ line.  While
addition of a broad iron line to the model significantly improves the
spectral fit, these residuals show that the evidence for the iron line
in these data is not very strong, probably because the observations
were shorter than the 1993 observation discussed by Eracleous, Halpern
\& Livio (1993).  Also, no evidence for a soft excess component is
seen.}
\end{figure}

The photon index and flux measured in the first 1995 {\it
ASCA} observation were consistent with those measured during the 1993
observation (Eracleous et al. \markcite{8} 1996).  The
column density is larger but only at the 68\% confidence level than
that measured by Eracleous et al. \markcite{8} (1996),
suggesting support for their idea that the absorption intrinsic to the
source changes slowly over time.  It also confirms their observation
that the absorption is significantly larger than the Galactic column
in that direction of $3.7 \times 10^{20}\rm\,cm^{-2}$ (Murphy et al.\
\markcite{20} 1996).  Analysis of a 20ks archival {\it ROSAT} PSPC
observation also shows absorption in excess of Galactic at $\sim
6\times 10^{20}\rm\,cm^{-2}$.

The recent reanalysis done by Wozniak et al.\ \markcite{23} (1996)
shows that there is evidence for a weak reflection component (e.g.
Nandra \& Pounds
\markcite{17} 1994) in the {\it Ginga} spectra from 3C~390.3.  In the
presence of a reflection component, spectral variability could be
observed if the relative normalization between the power-law and
reflection varied.  Physically, this could occur if the reflection
region is extended or very far from the nucleus, so that the change of
reflected flux is lagged or smeared compared with the input flux (e.g.
Leighly et al.\ \markcite{14} 1996).  In this case, the spectrum would
be expected to steepen as the flux increases, as observed.  However,
in the standard case, i.e. if the isotropic X-ray source is
illuminating cold material subtending $2\pi$ steradians, the
reflection component cannot be detected below about 5 keV (e.g.
Weaver et al.\ \markcite{22} 1995).  To test the effect of reflection
on the spectral variability, we fit the spectra simultaneously for
energies below 5 keV.  In Figure~3 the photon index vs $N_H$ $\chi^2$
contours for this fit are shown (dashed line).  These indicate that
the spectral variability can still be detected in this band,
demonstrating that it cannot be attributed to a change in the relative
normalization between the reflection and power-law components.

\begin{figure}[h]
\vbox to3.25in{\rule{0pt}{3.25in}}
\includegraphics{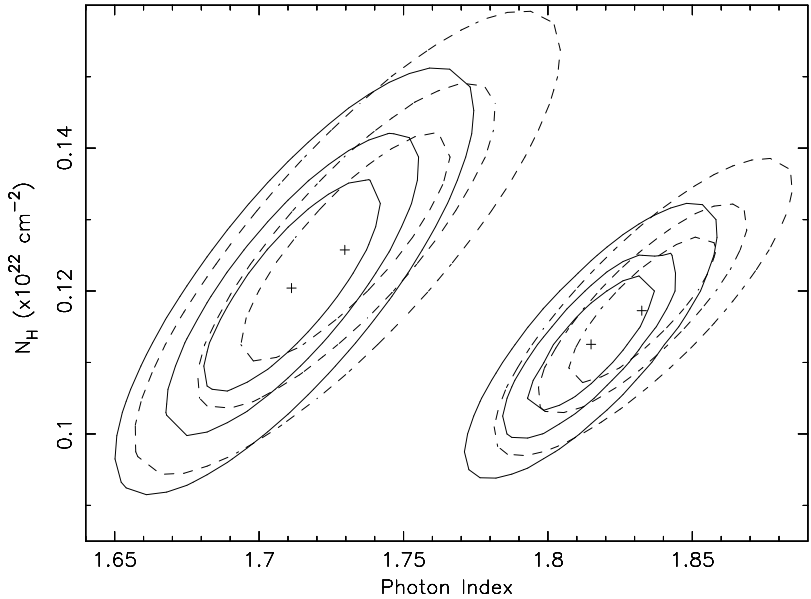}
\caption{Photon index versus $N_H$ $\chi^2$ contours (99\%, 90\%
and 68\%) for the two {\it ASCA} observations.  These demonstrate that
the photon index differs significantly between these two observations.
The solid line shows the results from fitting over the full 0.5--10.0
keV range, while the dashed line shows the results from the restricted
range 0.5--5 keV.  The results are consistent for the two ranges,
demonstrating that the spectral variability cannot be attributed to a
change in the relative normalization of the power-law and reflection
components.}
\end{figure}

\subsection{The Spectral Variability}

The photon index change in the {\it ASCA} data, with an increase in
flux, implies a pivot point for the spectrum of $\sim 400\rm \,keV$
for 3C~390.3.  Zdziarski et al.\ \markcite{25} (1995) discuss the
average X-ray and $\gamma$-ray spectra from {\it Ginga} and {\it CGRO}
OSSE from groups partitioned according to their optical and radio
characteristics.  The mean spectrum from the radio-loud group,
consisting of 3C~390.3 and 3C~311, when fit with a thermal
Comptonization model, gave a thermal cutoff of
$E_C=480^{+670}_{-190}\rm\,keV$.  While not well constrained, this
value is consistent with the pivot point of the spectral variability.

Photon index variability has been previously observed in the Ginga
spectra from 3C~390.3 (Inda et al.\ \markcite{10} 1991; Wozniak
\markcite{23} et al.\ 1996).  Table 4 gives a compilation of the photon
indices and fluxes from the Ginga and {\it ASCA} observations.  We
don't consider observations from previous observatories including
EXOSAT because the parameters are much more poorly constrained for
this fairly faint source (Inda et al.\ \markcite{10} 1991).  Because
we are interested in the intrinsic photon index, we use values from
fits which include the reflection model taken from Wozniak et al.\
\markcite{23} (1997) and Eracleous et al. \markcite{8} (1996).  For
the {\it ASCA} observations reported in this paper, we use the values
obtained from fits below $5\rm\,keV$.  The flux plotted against the
photon index is shown in Figure 4.

\begin{figure}[h]
\vbox to3.25in{\rule{0pt}{3.25in}}
\includegraphics{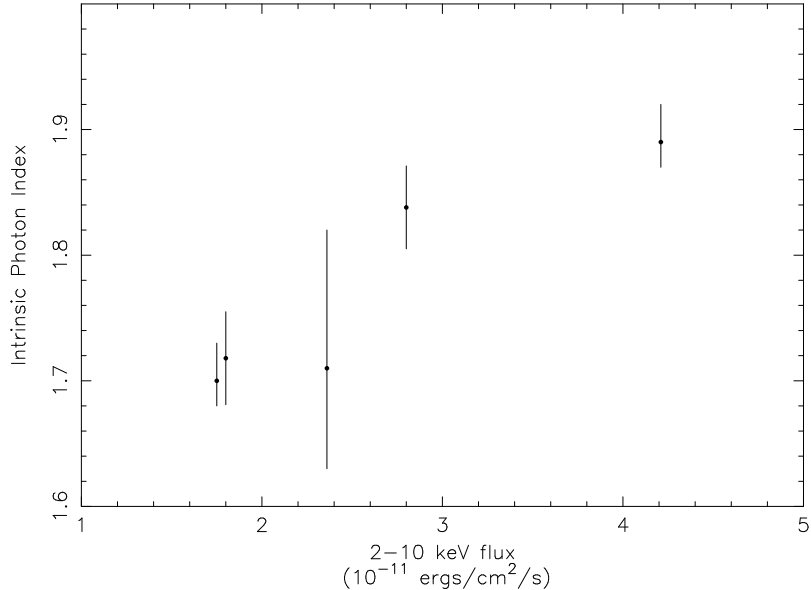}
\caption{Photon index versus $N_H$ $\chi^2$ contours (99\%, 90\%
and 68\%) for the two {\it ASCA} observations.  These demonstrate that
the photon index differs significantly between these two observations.
The solid line shows the results from fitting over the full 0.5--10.0
keV range, while the dashed line shows the results from the restricted
range 0.5--5 keV.  The results are consistent for the two ranges,
demonstrating that the spectral variability cannot be attributed to a
change in the relative normalization of the power-law and reflection
components.}
\end{figure}

The photon index changes from roughly 1.7 to 1.9 as the flux doubles,
and a general correlation of the flux and photon index is seen.  This
is similar to that observed from NGC~4151 (Yaqoob \& Warwick
\markcite{24} 1991) and NGC~5548 (Nandra \markcite{16}
et al.\ 1993).  Recently, Haardt, Maraschi \& Ghisellini \markcite{9}
(1997) investigated spectral variability in the context of thermal
disk-corona models.  The change in photon index measured here for the
factor of two change in flux is too large to be explained by the
simplest pair-dominated reprocessing models.  Rather, it suggests that
scattering optical depth of the corona is varying independently of the 
luminosity (Haardt et al. \markcite{9} 1997).

\section{Summary} 

In this paper we present the {\it ROSAT} and {\it ASCA} observations
obtained during the 1995 monitoring campaign of 3C~390.3.  The results
can be summarized as follows:
\begin{itemize}
\item We obtained the first well sampled X-ray light curve from an
AGN on time scales of days to months.  Large amplitude flaring was
observed with variability time scale of about 12 days, and quiecent
periods with markedly reduced variability were observed on time scales as long
as 30 days.  While the amplitude of variability is larger than
previously observed during X-ray monitoring campaigns, this is
probably due to the fact that this campaign was much longer.  While
the time scale of variability seems slower than in many other well
studied X-ray variable AGN, this may be reconciled with the greater
luminosity ($2-4\times 10^{44} \rm ergs\,cm^{-2}\,s^{-1}$).
\item The HRI hardness ratio varies during the monitoring campaign,
but this is likely due to the known time dependent gain change of the
instrument and therefore cannot be used to investigate the spectral
variability.
\item  The spectra from two {\it ASCA} observations made during the
monitoring period can be fit with an absorbed power law and an iron
$K\alpha$ line.  No evidence for soft excess emission was found,
indicating that the {\it ROSAT} HRI is sampling the X-ray power law.
The absorption is significantly larger than the Galactic value and
slightly larger than that found in a 1993 {\it ASCA} observation
(Eracleous et al. \markcite{8} 1996).  The photon index from the
brighter observation was significantly larger than that from the
fainter observation, and a compilation of results from {\it ASCA} and
{\it Ginga} observations indicate that the photon index is correlated
with the flux on long time scales, a trend seen in several Seyfert 1
galaxies.
\end{itemize}

\acknowledgements

K.M.L. thanks Dan Harris and Andrea Prestwich for useful discussions
regarding the {\it ROSAT} HRI calibration, and Tahir Yaqoob for a
critical reading of the manuscript.  K.M.L and R.E. acknowledge
support through {\it ASCA} AO-3 Guest Observers program grant
NAG~5-2547 and {\it ROSAT} AO-5 Guest Observers program grant
NAG~5-2637.  K.M.L. acknowledges support at RIKEN by a Science and
Technology Agency fellowship.

\clearpage

\begin{deluxetable}{lrrr}
\small
\tablewidth{0pc}
\tablenum{1}
\tablecaption{{\it ROSAT} HRI Observation Log and Fluxes}
\tablehead{
\colhead{Sequence Number}   & \colhead{Julian Date\tablenotemark{1}} & 
\colhead{Exposure\tablenotemark{2}} & \colhead{Count Rate\tablenotemark{3}}  \\
 & \colhead{(2,440,000+)} & \colhead{(s)} & \colhead{($\rm counts\,s^{-1}$)}}

\startdata

rh701798 & 9719.694 & 2927 & $0.301\pm0.011$ \nl
rh701799 & 9722.684 & 2098 & $0.454\pm0.016$ \nl
rh701800 & 9725.538 & 1825 & $0.485\pm0.017$ \nl
rh701801 & 9728.656 & 1673 & $0.421\pm0.017$ \nl
rh701802 & 9731.576 & 1353 & $0.331\pm0.017$ \nl
rh701803 & 9734.561 & 1305 & $0.321\pm0.017$ \nl
rh701804 & 9737.546 & 1052 & $0.380\pm0.020$ \nl
rh701805 & 9740.597 & 1043 & $0.447\pm0.021$ \nl
rh701806 & 9744.312 & 712 & $0.384\pm0.024$ \nl
rh701807 & 9747.350 & 2941 & $0.349\pm0.012$ \nl
rh701808 & 9750.200 & 2999 & $0.436\pm0.013$ \nl
rh701809 & 9752.588 & 2521 & $0.424\pm0.014$ \nl
rh701810 & 9755.577 & 1588 & $0.362\pm0.016$ \nl
rh701811 & 9758.565 & 1253 & $0.354\pm0.018$ \nl
rh701812 & 9761.948 & 3159 & $0.394\pm0.012$ \nl
rh701813 & 9764.941 & 1788 & $0.353\pm0.015$ \nl
rh701814 & 9767.595 & 2190 & $0.321\pm0.013$ \nl
rh701815 & 9770.547 & 1259 & $0.336\pm0.017$ \nl
rh701816 & 9773.618 & 2315 & $0.340\pm0.013$ \nl
rh701817 & 9776.582 & 1268 & $0.321\pm0.017$ \nl
rh701818 & 9778.909 & 2045 & $0.333\pm0.014$ \nl
rh701819 & 9781.763 & 2618 & $0.374\pm0.013$ \nl
rh701820 & 9784.685 & 2578 & $0.457\pm0.014$ \nl
rh701821 & 9787.804 & 2219 & $0.588\pm0.017$ \nl
rh701822 & 9790.926 & 1311 & $0.778\pm0.025$ \nl
rh701823 & 9793.642 & 1834 & $0.964\pm0.024$ \nl
rh701824 & 9796.561 & 1794 & $0.844\pm0.022$ \nl
rh701825 & 9799.980 & 1997 & $0.685\pm0.019$ \nl
rh701826 & 9802.731 & 2287 & $0.544\pm0.016$ \nl
rh701827 & 9805.583 & 1473 & $0.547\pm0.020$ \nl
rh701796 & 9809.564 & 1316 & $0.560\pm0.021$ \nl
rh701795 & 9812.538 & 2922 & $0.558\pm0.015$ \nl
rh701794 & 9815.595 & 1896 & $0.530\pm0.018$ \nl
rh701793 & 9818.647 & 1592 & $0.514\pm0.019$ \nl
rh701730 & 9819.943 & 2430 & $0.534\pm0.016$ \nl
rh701731 & 9823.554 & 1641 & $0.567\pm0.020$ \nl
rh701732 & 9825.544 & 1551 & $0.497\pm0.019$ \nl
rh701733 & 9830.451 & 1460 & $0.529\pm0.020$ \nl
rh701734 & 9831.568 & 1800 & $0.484\pm0.018$ \nl
rh701735 & 9834.585 & 1483 & $0.452\pm0.019$ \nl
rh701737 & 9840.586 & 2815 & $0.548\pm0.015$ \nl
rh701738 & 9843.570 & 3048 & $0.594\pm0.015$ \nl
rh701739 & 9846.590 & 2224 & $0.530\pm0.016$ \nl
rh701740 & 9849.673 & 1531 & $0.483\pm0.019$ \nl
rh701742 & 9855.577 & 1906 & $0.428\pm0.016$ \nl
rh701743 & 9858.563 & 1586 & $0.488\pm0.019$ \nl
rh701744 & 9861.586 & 1394 & $0.596\pm0.021$ \nl
rh701745 & 9864.768 & 2030 & $0.557\pm0.017$ \nl
rh701746 & 9867.950 & 2120 & $0.604\pm0.018$ \nl
rh701747 & 9870.869 & 2119 & $0.602\pm0.018$ \nl
rh701748 & 9873.853 & 2037 & $0.530\pm0.017$ \nl
rh701749 & 9876.838 & 1717 & $0.507\pm0.018$ \nl
rh701750 & 9879.711 & 2457 & $0.536\pm0.016$ \nl
rh701751 & 9882.662 & 3209 & $0.446\pm0.013$ \nl
rh701752 & 9885.624 & 1168 & $0.425\pm0.020$ \nl
rh701753 & 9888.599 & 1910 & $0.414\pm0.015$ \nl
rh701754 & 9891.613 & 1663 & $0.489\pm0.018$ \nl
rh701755 & 9894.567 & 1687 & $0.397\pm0.016$ \nl
rh701757 & 9900.600 & 1401 & $0.266\pm0.015$ \nl
rh701758 & 9903.550 & 1883 & $0.255\pm0.013$ \nl
rh701759 & 9906.568 & 1891 & $0.289\pm0.013$ \nl
rh701760 & 9909.586 & 1538 & $0.300\pm0.015$ \nl
rh701761 & 9912.670 & 2675 & $0.308\pm0.012$ \nl
rh701762 & 9915.651 & 1805 & $0.316\pm0.015$ \nl
rh701763 & 9918.544 & 1187 & $0.333\pm0.018$ \nl
rh701764 & 9921.593 & 3089 & $0.354\pm0.012$ \nl
rh701765 & 9924.778 & 2137 & $0.352\pm0.014$ \nl
rh701766 & 9927.797 & 1639 & $0.321\pm0.015$ \nl
rh701767 & 9930.780 & 1804 & $0.391\pm0.016$ \nl
rh701768 & 9933.930 & 1629 & $0.399\pm0.017$ \nl
rh701769 & 9936.980 & 1562 & $0.511\pm0.019$ \nl
rh701770 & 9940.131 & 2318 & $0.507\pm0.015$ \nl
rh701772 & 9945.732 & 2051 & $0.633\pm0.018$ \nl
rh701773 & 9948.718 & 1885 & $0.624\pm0.019$ \nl
rh701774 & 9951.635 & 1975 & $0.660\pm0.019$ \nl
rh701775 & 9954.619 & 2065 & $0.791\pm0.020$ \nl
rh701776 & 9957.603 & 1943 & $0.894\pm0.022$ \nl
rh701777 & 9960.587 & 1805 & $0.855\pm0.022$ \nl
rh701778 & 9963.639 & 2550 & $0.733\pm0.018$ \nl
rh701779 & 9966.578 & 1595 & $0.664\pm0.021$ \nl
rh701780 & 9969.563 & 1925 & $0.704\pm0.020$ \nl
rh701781 & 9972.580 & 1847 & $0.967\pm0.023$ \nl
rh701782 & 9975.643 & 1956 & $0.862\pm0.021$ \nl
rh701783 & 9978.583 & 1689 & $0.731\pm0.021$ \nl
rh701784 & 9981.531 & 2585 & $0.688\pm0.017$ \nl
rh701785 & 9984.582 & 1336 & $0.558\pm0.021$ \nl
rh701786 & 9987.606 & 1847 & $0.571\pm0.018$ \nl
rh701787 & 9990.625 & 2527 & $0.636\pm0.016$ \nl
rh701788 & 9993.644 & 933 & $0.790\pm0.030$ \nl
rh701789 & 9996.761 & 1764 & $0.877\pm0.023$ \nl
\enddata
\tablecomments{Exposures and count rates have not been corrected for
the off-axis response.}
\tablenotetext{1}{Observation time is given for midpoint of each
observation; no barycentric correction has been performed.}
\tablenotetext{2}{Exposure has been corrected for livetime.}
\tablenotetext{3}{Using the distributed HRI response matrix and ignoring
the effects of the gain change, 1~count/s corresponds to 2.93 and
$3.04 \times 10^{-11}\rm\,ergs\,s^{-1}cm^{-2}$ for photon indices 1.7
and 1.82 respectively and $N_H=1.15\times 10^{21}\rm\, cm^{-2}$.}

\end{deluxetable}

\clearpage

\begin{deluxetable}{lcc}
\tablewidth{0pc}
\tablenum{2}
\tablecaption{{\it ASCA} Observation Log}
\tablehead{
\colhead{Detector}   & \colhead{Exposure} & 
\colhead{Net Count Rate} \\
& & \colhead{($\rm counts\,s^{-1}$)}}

\startdata

\multicolumn{3}{l}{First Observation (TJD 9733.12)\tablenotemark{a}} \nl
SIS0 & 18204 & 0.604 \nl
SIS1\tablenotemark{b} & 10844 & 0.509 \nl
GIS2 & 18122 & 0.356 \nl
GIS3 & 18122 & 0.432 \nl
\tableline
\multicolumn{3}{l}{Second Observation (TJD 9843.75)\tablenotemark{a}} \nl
SIS0 & 17298 & 1.080 \nl
SIS1 & 17300 & 0.899 \nl
GIS2 & 18760 & 0.600 \nl
GIS3 & 18770 & 0.702 \nl
\tablenotetext{a}{Julian date of midpoint of the observation, 2,440,000+.}
\tablenotetext{b}{The exposure time for this detector is lower,
because only the data collected at high bit rate were used for
spectral analysis; see text for details.}
\enddata
\end{deluxetable}

\begin{deluxetable}{lcc}
\small
\tablewidth{0pc}
\tablenum{3}
\tablecaption{{\it ASCA} Spectral Fitting Results}
\tablehead{
\colhead{Parameter}   & \colhead{First Observation}
   & \colhead{Second Observation}}

\startdata

$N_H (\times 10^{22}\rm cm^{-2})$ & \multicolumn{2}{c}
{$0.115_{-0.011}^{+0.012}$}  \nl
Photon Index & $1.70 \pm 0.03$ & $1.82 \pm 0.03$ \nl 
SIS0 PL norm ($\times 10^{-3} \rm
photons\,keV^{-1}cm^{-2}s^{-1}$) & $4.39 \pm 0.15$ &
$8.18^{+0.26}_{-0.25}$ \nl
Iron Line Energy (keV) & \multicolumn{2}{c}
{$6.29^{+0.27}_{-0.28}$}  \nl
Iron Line Width (keV) & \multicolumn{2}{c}
{0.34 (fixed)}  \nl
Iron Line norm ($\times 10^{-5}\rm\,photons\,cm^{-2}\,s^{-1}$) &
$3.2_{-2.4}^{+2.3}$ &
$3.5_{-2.6}^{+2.9}$ \nl
Equivalent Width (eV) & $167\pm 120$ & $120_{-90}^{+100}$ \nl
Observed Flux (2--10 keV; $\times 10^{-11}\rm
\,ergs\,cm^{-2}s^{-1}$) & $1.8$ & $2.8$ \nl
Intrinsic Luminosity (2--10 keV; $\times 10^{44}\,\rm ergs\,s^{-1}$) &
2.6 & 4.0 \nl 
\tablecomments{The best fit $\chi^2=2165/2129$ d.o.f. All quoted errors are 90\% confidence for two
parameters of interest ($\Delta\chi^2=4.61$).  The line width could
not be constrained, so it was fixed at the best fit value during
evaluation of errors of the other parameters.}

\enddata
\end{deluxetable}

\begin{deluxetable}{llccc}
\tablewidth{0pc}
\tablenum{4}
\tablecaption{Photon Index History}
\tablehead{
\colhead{Date}   & \colhead{Instrument} & 
\colhead{2-10 keV Flux} & \colhead{Photon Index} &
\colhead{Reference} \\
& & \colhead{($\times 10^{-11}\rm\,ergs\,cm^{-2}\,s^{-1}$)} & & }

\startdata

1988.86 & {\it Ginga} & 4.2 & 
\tablenotemark{a}\hphantom{h}$1.89^{+0.03}_{-0.02}$ & 1 \nl
1991.12 & {\it Ginga} & 2.4 & 
\tablenotemark{a}\hphantom{h}$1.71^{+0.11}_{-0.08}$ & 1 \nl
1993.88 & {\it ASCA} SIS & 1.75 & 
\tablenotemark{a}\hphantom{h}$1.70^{+0.03}_{-0.02}$ & 2 \nl
1995.15 & {\it ASCA} SIS+GIS & 1.8 & 
\tablenotemark{b}\hphantom{h}$1.72\pm 0.04$ &  3 \nl
1995.125 &  {\it ASCA} SIS+GIS & 2.8 & 
\tablenotemark{b}\hphantom{h}$1.84\pm 0.03$ & 3 \nl
\tablerefs{(1) Wozniak et al.\ 1996; (2) Eracleous et al. 1996; 
(3) this paper.}
\tablenotetext{a}{Obtained from fits which use a reflection model (see text).}
\tablenotetext{b}{Obtained from fitting spectra below 5 keV only (see text).}

\enddata
\end{deluxetable}


\clearpage

\begin{references}

\reference{1} Alef, W., Wu, S. Y., Preuss, E., Kellerman, K. I., \&
Qiu, Y. H. 1996, \aap, 308, 376

\reference{2} Barr, P., et al.\ 1980, \mnras, 193, 549

\reference{3} Clavel, J., \& Wamsteker, W. 1987, ApJ, 329, 9

\reference{4} David, L. P., Harnden, F. R., Kearns, K. E., \& Zombeck,
M. V.\ 1996, {\it The ROSAT High Resolution Imager (HRI) Calibration Report}

\reference{5} Dietrich, M., et al.\ 1997, in preparation

\reference{6} Edelson, R., et al.\ 1996, \apj, 470, 364

\reference{7} Eracleous, M. \& Halpern, J. P. 1994, \apjs, 90, 1

\reference{8} Eracleous, M., Halpern, J. P., \& Livio, M. 1996, \apj,
459, 89

\reference{9} Haardt, F., Maraschi, L., \& Ghisellini, G. 1997, ApJ,
in press

\reference{10} Inda, M., et al.\ 1994, \apj, 420, 143

\reference{11} Lawrence, A., \&  Papadakis, I. E., 1993 ApJL, 414, 85

\reference{14} Leighly, K. M., Kunieda, H., Awaki, H., \& Tsuruta, S.
1996, \apj, 463, 158

\reference{12} Leighly, K. M., \& O'Brien, P. T.  1997, accepted for
publication in ApJL

\reference{13} Leighly, K. M., et al.  1997 in preparation

\reference{26} Matt, G., Perola, G. C., Piro, L., \& Stella, L. 1992,
\aap, 257, 63

\reference{20} Murphy, E. M., Lockman, F. J., Laor, A., \& Elvis, M.
1996, \apjs, 105, 369

\reference{15} Mushotzky, R. F., Baity, W. A., \& Peterson, L. E. 1977,
\apj, 212, 22

\reference{16} Nandra, K., Pounds, K. A., Stewart, G. C., George, I.
M., \& Hayashida, K. 1991, \mnras, 248, 760

\reference{17} Nandra, K., \& Pounds, K. A. 1994, \mnras, 268, 405

\reference{18} O'Brien, P. T., et al.\ 1997, in preparation

\reference{19} Reynolds, C. S., Fabian, A. C., Nandra, K., Inoue, H.,
Kunieda, H., \& Iwasawa, K. 1995, \mnras, 277, 901

\reference{21} Veilleux, S. \& Zheng, W. 1991, \apj, 377, 89

\reference{22} Weaver, K. A., Nousek, J., Yaqoob, T., Hayashida, K. \&
Murakami, S. 1995, \apj, 451, 147

\reference{23} Wozniak, P.R., Zdziarski A.A., Smith D., Madejski G.M. 1997,
submitted to MNRAS

\reference{24} Yaqoob, T., \& Warwick, R. S. 1991, \mnras, 248, 773

\reference{25} Zdziarski, A. A., Johnson, W. N., Done, C., Smith, D.,
\& McNaron-Brown, K. 1995, ApJL

\end{references}
\end{document}